\documentclass[%
 preprint,
 superscriptaddress,
 amsmath,amssymb,
 aps,
 showkeys,
 titlepage,
 endfloats*.   
]{revtex4-2}

\usepackage[utf8]{inputenc}
\usepackage[english]{babel}

\usepackage{graphicx}
\usepackage{dcolumn}
\usepackage{bm}

\usepackage[colorlinks = true,
            linkcolor = blue,
            urlcolor  = blue,
            citecolor = red,
            anchorcolor = blue]{hyperref}

\begin{document}


\title{Imaging Hot Photocarrier Transfer across a Semiconductor Heterojunction with Ultrafast Electron Microscopy} 

\author{Basamat S. Shaheen}
\affiliation{Department of Mechanical Engineering, University of California, Santa Barbara, CA 93106, USA}

\author{Kenny Huynh}
\affiliation{Department of Materials Science and Engineering, University of California, Los Angeles, CA 90095, USA}

\author{Yujie Quan}
\affiliation{Department of Mechanical Engineering, University of California, Santa Barbara, CA 93106, USA}

\author{Usama Choudhry}
\affiliation{Department of Mechanical Engineering, University of California, Santa Barbara, CA 93106, USA}

\author{Ryan Gnabasik}
\affiliation{Department of Mechanical Engineering, University of California, Santa Barbara, CA 93106, USA}

\author{Zeyu Xiang}
\affiliation{Department of Mechanical Engineering, University of California, Santa Barbara, CA 93106, USA}

\author{Mark Goorsky}
\email{goorsky@seas.ucla.edu}
\affiliation{Department of Materials Science and Engineering, University of California, Los Angeles, CA 90095, USA}

\author{Bolin Liao}
\email{bliao@ucsb.edu} \affiliation{Department of Mechanical Engineering, University of California, Santa Barbara, CA 93106, USA}

\date{\today}

\begin{abstract}
Semiconductor heterojunctions have gained significant attention for efficient optoelectronic de-
vices owing to their unique interfaces and synergistic effects. Interaction between charge carriers with the heterojunction plays a crucial role in determining device performance, while its spatial-temporal mapping remains lacking. In this study, we employ scanning ultrafast electron microscopy (SUEM), an emerging technique that combines high spatial-temporal resolution and surface sensitivity, to investigate photocarrier dynamics across a Si/Ge heterojunction. Charge dynamics are selectively examined across the junction and compared to far bulk areas, through which the impact of the built-in potential, band offsets, and surface effects is directly visualized. In particular, we find that the heterojunction drastically modifies the hot photocarrier diffusivities by up to 300\%. These findings are further elucidated with insights from the band structure and surface potential measured by complementary techniques. This work demonstrates the tremendous effect of hetero-interfaces on charge dynamics and showcases the potential of SUEM in characterizing realistic devices.

\end{abstract}

                            
\maketitle


\section*{One Sentence Summary}
Heterojunction drastically modifies diffusion of photoexcited hot charge as directly visualized by short electron pulses. 

\section{Introduction}


In many solid-state device applications, the creation of junctions within semiconductor materials provides a powerful approach to overcome the limitations of individual materials \cite{essig2017raising}.  
Semiconductor junctions come in two types: homojunctions and heterojunctions. A homojunction occurs within a single semiconductor material with varying doping levels, whereas a heterojunction forms when two distinct semiconductor materials come into contact with each other \cite{sze2008semiconductor}. 
While p-n homojunctions are fundamental in semiconductor devices, the bulk of research attention is shifted towards heterojunctions where each side of the junction is made of a distinct semiconductor material with different compositions, bandgaps, crystal structures, or lattice constants \cite{faber2020heterojunction,aydin2024pathways,liu2021boosting,giannazzo2019engineering,oh2017double,wu2021designs}.
This diversity translates into a broader material spectrum, unlocking complementary properties and synergistic effects to achieve superior device performance and opening up opportunities for innovative device concepts and applications across various fields. 


The interface plays a crucial role in the functionality of a heterojunction device, far beyond serving as a divider between a homostructure device and a chemically dissimilar substrate \cite{ferry1991heterojunctions}. Often, the interface itself becomes the operational core of the device, shaping its characteristics and performance \cite{kroemer1983heterostructure}. Challenges in designing high-performance heterostructures, as outlined by Kroemer, revolve around understanding the energy band structure and elucidating charge transport phenomena across the interface \cite{kroemer1983heterostructure}. In particular, how energetic charge carriers injected either electrically or via photoexcitation interact with the heterojunction is the central factor that controls the device performance. Theoretical models can provide a starting point to estimate energy band offsets and built-in potentials for a heterojunction and predict charge transport behaviors accordingly~\cite{margaritondo2012electronic,sharma2015semiconductor,Zhang2012BandBI}. However, it has been a challenge to accurately simulate charge carrier dynamics in a mesoscopic structure, such as a heterojunction, where nonidealities associated with defects induced by processing limit the theoretical predictive power. Fundamentally, transport of highly non-equilibrium ``hot'' carriers across a heterojunction has not been fully modeled and understood~\cite{lin2019asymmetric}. Experimentally, the influence of the heterojunction on charge transport is often inferred indirectly from electrical transport measurements~\cite{nourbakhsh2016transport}, lacking the resolution to correlate local structures and potentials with charge behaviors. 
For these reasons, there exists a critical need for direct surface characterization techniques with sufficient space and time resolutions tailored to device-type structures, capable of yielding an accurate understanding of interface physics parameters and the complex charge dynamics near heterojunctions \cite{monch2013semiconductor,grundmann2010physics,xue2021interfacial,luo2021recent}.

Time-resolved spectroscopy has been widely used to study charge transfer processes in heterojunctions occuring on an ultrafast time scale \cite{mitchell2021interfacial,spies2023time,yang2015semiconductor,jakowetz2017visualizing,zhu2017interfacial}. 
Recently, transient microscopy techniques have been employed with high spatial-temporal resolution to observe charge dynamics localized across interfaces \cite{yue2022high,davydova2016transient,vazquez2024spatiotemporal,walsh2024monolayer}. 
While these methods provide valuable  information, the large penetration depth of the optical probe obtains dynamical information mainly from the bulk \cite{zewail2010four}. 
Surface-sensitive techniques such as X-ray photoelectron spectroscopy (XPS) and scanning Kelvin probe microscopy (SKPM) are capable of providing chemical information and high-resolution imaging of surface potential variations across interfaces, respectively \cite{chambers2020introductory,morgan2022band,chen2017insight,sharma2017enhanced}. However, they usually lack the time resolution to directly map the charge dynamics on relevant time scales. Recently, time-resolved photoemission electron microscopy (TR-PEEM) has been used to probe charge transfer across a two-dimensional hetero-interface with high temporal and spatial resolutions~\cite{man2017imaging}.
Combining high spatial-temporal resolutions and high sensitivity to surface charge carrier dynamics, scanning ultrafast electron microscopy (SUEM) has emerged as a suitable tabletop method to visualize charge transfer across interfaces ~\cite{yang2010scanning,liao2017scanning}. It utilizes a pulsed electron probe to scan the surface, eliciting secondary electrons (SEs) from the top few nanometers of materials (typically less than 10\,nm). 
As a result, the photoinduced contrast images obtained through SUEM, reflecting local changes in SE emission, capture dynamic information exclusively from the surface ~\cite{adhikari2017four}. SUEM has been successfully applied to directly image photocarrier dynamics in a wide range of materials~\cite{liao2017photo,liao2017spatial,kim2021transient,choudhry2023persistent,El-Zohry2019extra}.
Previously, SUEM was applied to study carrier dynamics in a silicon p-n homojunction~\cite{najafi2015four} and a MoS$_2$ homojunction~\cite{wong2021spatiotemporal}, where the charge separation effect by the built-in potential at a homojunction was examined. Compared to a homojunction, heterojunctions can host more complex charge carrier dynamics due to the additional band offsets and different surface conditions on dissimilar materials forming the junction, whose impact on photocarrier transport remains less explored.  

In this work, we apply SUEM to provide a holistic view of photoexcited charge dynamics in a Si/Ge heterojunction. 
The Si/Ge heterojunction samples fabricated through wafer bonding were chosen as our test model for this study due to the extensive research on both constituents and the wide range of applications for their heterojunctions ~\cite{haddara2017silicon,basu2022review}. 
By comparing SUEM contrast images taken at the heterojunction to those taken at bulk Si and Ge regions, we directly visualize the significant influence of the junction on photocarrier dynamics. In particular, we find that the built-in potential and the band offsets drastically change the hot photocarrier diffusivities on both Si and Ge side.
Integrating our findings from SUEM with systematic XPS and SKPM characterizations allows us to draw a complete picture of photoexcited charge transport across the interface. 
Through this experimental surface characterization approach, we aim to achieve a better grasp and control of device interfaces, driving forward advances in semiconductor technology.

\section{Results and Discussions}

\subsection{Quantifying the Heterojunction Parameters}
\begin{figure}[!h]
\includegraphics[width=0.9\textwidth]{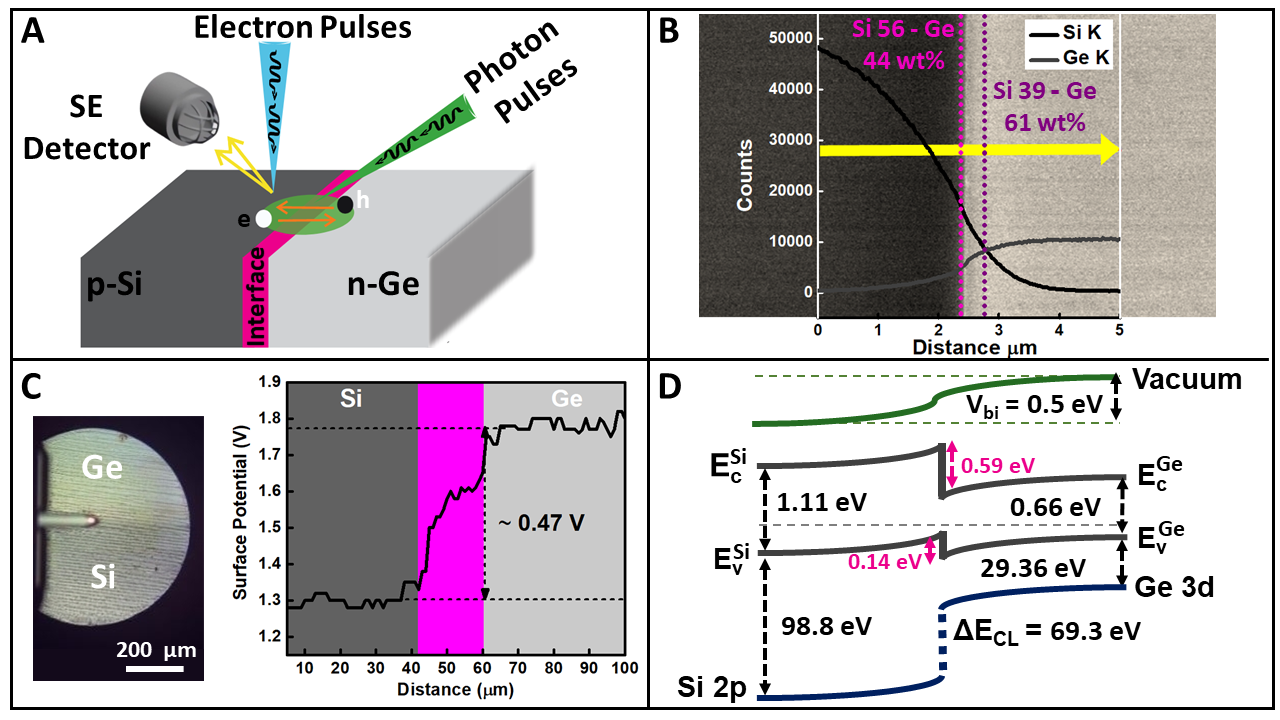}
\caption{\textbf{Schematic of the SUEM experiment and other characterizations of the sample.} (a) Schematic illustration of the SUEM measurement of a Si/Ge heterojunction. SE: secondary electrons. (b) EDS mapping of the elemental composition across the junction interface. The X-ray counts corresponding to the Si $K$ line and Ge $K$ line are plotted against the position across the junction. The background is an SE contrast image taken in the same SEM prior to the EDS scan. Si region has a darker SE contrast compared to the Ge region. (c) Results of the scanning Kelvin probe microscopy across the Si/Ge heterojunction interface. The left panel shows an optical image of the sample with the scanning probe. (d) Surface band diagram of the Si/Ge heterojunction determined from the XPS and SKPM measurements. $E_C$ and $E_V$ denote the conduction band edge and the valence band edge, respectively. } 
\label{fig:fig1}
\end{figure}

A schematic of our SUEM measurement is illustrated in Fig.~\ref{fig:fig1}a. More details are provided in Materials and Methods. Briefly, a cross section of a Si/Ge heterojunction sample is examined using SUEM, which is an optical-pump-electron-probe technique. Optical excitations are generated in different regions of the heterojunction by an optical pump pulse (wavelength 505\,nm, pulse duration 200\,fs, beam diameter 30\,$\mu$m). The optical response of the sample is monitored subsequently by a delayed electron pulse (30 keV kinetic energy, 30 to 40 electrons per pulse, pulse duration of a few picoseconds). The response is recorded as contrast images of SE emission from different locations of the sample surface.

Before the SUEM measurement, we conduct extensive characterizations of the Si/Ge heterojunction sample to gain necessary information to help interpret the SUEM results. More details are provided in Materials and Methods. A transmission electron microscope (TEM) image of the interface is shown in Fig.~S2, indicating an atomically smooth interface after bonding. Figure~\ref{fig:fig1}b shows the elemental mapping based on energy dispersive spectroscopy (EDS) taken in a scanning electron microscope (SEM). X-ray counts corresponding to Si $K$ line and Ge $K$ line are measured across the junction. The elemental mapping suggests interdiffusion of Si and Ge atoms across the interface during the bonding process, making the Si/Ge interface a graded junction over a range of a few micrometers. Therefore, it is expected that the built-in potential ($V_{bi}$) developed near the junction as a result of charge transfer will be spread across the lengthscale of a few micrometers. The background of Fig.~\ref{fig:fig1}b shows a representative SE contrast image of the Si/Ge heterojunction, where the Ge region exhibits a brighter contrast, signaling a higher SE yield. Given the primary beam energy at 30\,keV, this difference in SE yield is likely due to the heavier atomic mass of Ge and the resulted larger elastic scattering cross section~\cite{reimer1980measuring}.

To determine the crucial electrical properties of the heterojunction, including band offsets, built-in potential, and work functions, we further examined the sample using scanning Kelvin probe microscopy (SKPM) and X-ray photoemission spectroscopy (XPS). Both techniques, together with SUEM, are highly sensitive to the sample surface (top few nanometers) and, thus, reflect the junction properties as affected by the surface conditions, including surface band bending~\cite{li2020probing}. SKPM measures the contact potential difference (CPD) between the sample surface and the scanning tip, which is determined by the difference in their work functions~\cite{melitz2011kelvin}. Since the same scanning tip is used for measurement across the heterojunction, the corresponding CPD difference indicates the evolution of the work function across the junction. As shown in Fig.~\ref{fig:fig1}c, a higher CPD on the Ge side ($\sim$0.5\,eV higher than that on the Si side on average) is measured, suggesting a higher work function on the Ge side and a built-in potential across the junction $V_{bi} \approx$ 0.5\,eV. In addition, a smooth transition of the work function over a range of around 20 micrometers near the interface also suggests a graded junction, which is consistent with the EDS elemental mapping. XPS detects the binding energy of valence and core electrons as referenced from the Fermi level. XPS measurements of the relative positions between the valence band maximum (VBM) and specific core levels across a heterojunction can be used to estimate the band offsets at the junction~\cite{kraut1980precise,kraut1983semiconductor}. Raw XPS spectra and the associated analysis is provided in Fig.~S1 and Supplementary Note 1. From the analysis, the valence band offset between Si and Ge is determined to be 0.14\,eV and the conduction band offset 0.59\,eV. The complete surface band diagram of our Si/Ge heterojunction sample as determined by SKPM and XPS analyses is shown in Fig.~\ref{fig:fig1}d. We note here that the band diagram shown in Fig.~\ref{fig:fig1}d is based on surface-sensitive characterizations, where surface conditions can lead to electrical potential changes that can shift the band alignment. Surface effects can lead to discrepancies with band diagrams obtained via bulk transport measurements~\cite{gity2013ge}. Since SUEM is also a surface-sensitive technique, the surface band diagram shown in Fig.~\ref{fig:fig1}d is more relevant for interpretting SUEM results.  

\subsection{Understanding SUEM Contrast in the Bulk Regions}
\begin{figure}[!h]
\includegraphics[width=\textwidth]{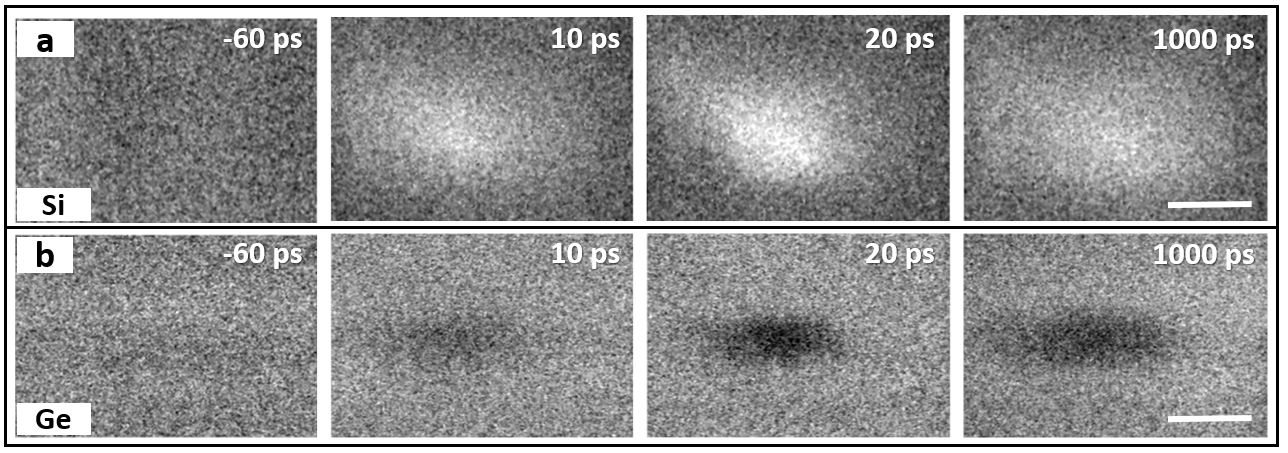}
\caption{\textbf{SUEM contrast images taken at bulk Si and Ge regions.} (a) SUEM contrast images taken at bulk Si region far away from the junction. (b) SUEM contrast images taken at bulk Ge region far away from the junction. The time label indicates the delay time between the optical pump pulse and the electronic probe pulse. Scale bar: 50 $\mu$m.} 
\label{fig:fig2}
\end{figure}

SUEM contrast images taken in Si and Ge bulk regions far away from the heterojunction are shown in Fig.~\ref{fig:fig2}. The time labels represent the delay time between the optical pump pulse and the electronic probe pulse. Negative times indicate the probe pulse arrives before the pump pulse. These contrast images are generated by subtracting a reference SUEM image taken at far negative times (-300 ps in our case) from SUEM images taken at later delay times. Therefore, the bright or dark contrasts seen in the SUEM images represent an increase or decrease in the SE yield as a result of the optical excitation. The lack of contrast at negative delay times in both Si and Ge bulk regions confirms that the sample has sufficient time to relax to its original state between two consecutive optical pump pulses (the pulses are separated by 200\,ns).

We observe bright SUEM contrast in Si bulk regions at positive delay times, suggesting an increased SE yield as a result of the optical excitation. The SUEM contrast mechanism in Si has been extensively investigated before~\cite{najafi2015four,najafi2017super,li2020probing,ouyang2023impact}. In heavily-doped Si, where surface band bending due to Fermi level pinning can significantly affect SE emission, photo-induced modification of the surface bands, the so called ``surface photovoltage'' effect, can give rise to bright (dark) contrast in n-type (p-type) Si~\cite{li2020probing,ouyang2023impact}. In lightly doped (as in this study) or intrinsic Si, in contrast, optical excitation of photocarriers raises the average electron energy in the bulk, which leads to a higher SE yield and bright contrast~\cite{mohammed20114D}. Therefore, the intensity of the bright contrast can be correlated to the local surface photocarrier concentration. Subsequent evolution of the bright contrast can be fitted to a Gaussian function, from which both the intensity and spatial extent of the photocarrier distribution can be extracted. The extracted intensity as a function of delay time in the Si bulk region is shown in Fig.~S3a. The initial rise of the intensity ($\sim$7\,ps) reflects the time resolution of the instrument while the later decay is due to the photocarrier recombination. In bulk Si region, the fitted recombination time ($\sim$2\,ns) agrees well with the radiative recombination time in Si. The evolution of the spatial distribution of the bright contrast represents the diffusion process of the photocarriers and will be analyzed later. Unlike Si, the Ge bulk region shows a dark contrast in the SUEM images, indicating a reduced SE yield as a result of photoexcitation. This behavior has been previously observed in GaAs~\cite{cho2014visualization}, where the dark contrast was attributed to increased scattering of the SEs by the photoexcited carriers. Alternatively, the dark contrast could be caused by the surface band bending effect that can occur on Ge surfaces that are less sensitive to doping concentrations~\cite{gobeli1964photoelectric}. In either case, the dark contrast can be correlated with the spatial distribution of photocarriers in Ge. The intensity of the Gaussian fit of the SUEM contrast is given in Fig.~S3b. In the bulk Ge region, the much shorter recombination time (141\,ps) suggests significant surface recombination potentially induced by surface defects.    

\subsection{Imaging charge transfer across Si/Ge heterojunction}

\begin{figure}[!h]
\includegraphics[width=\textwidth]{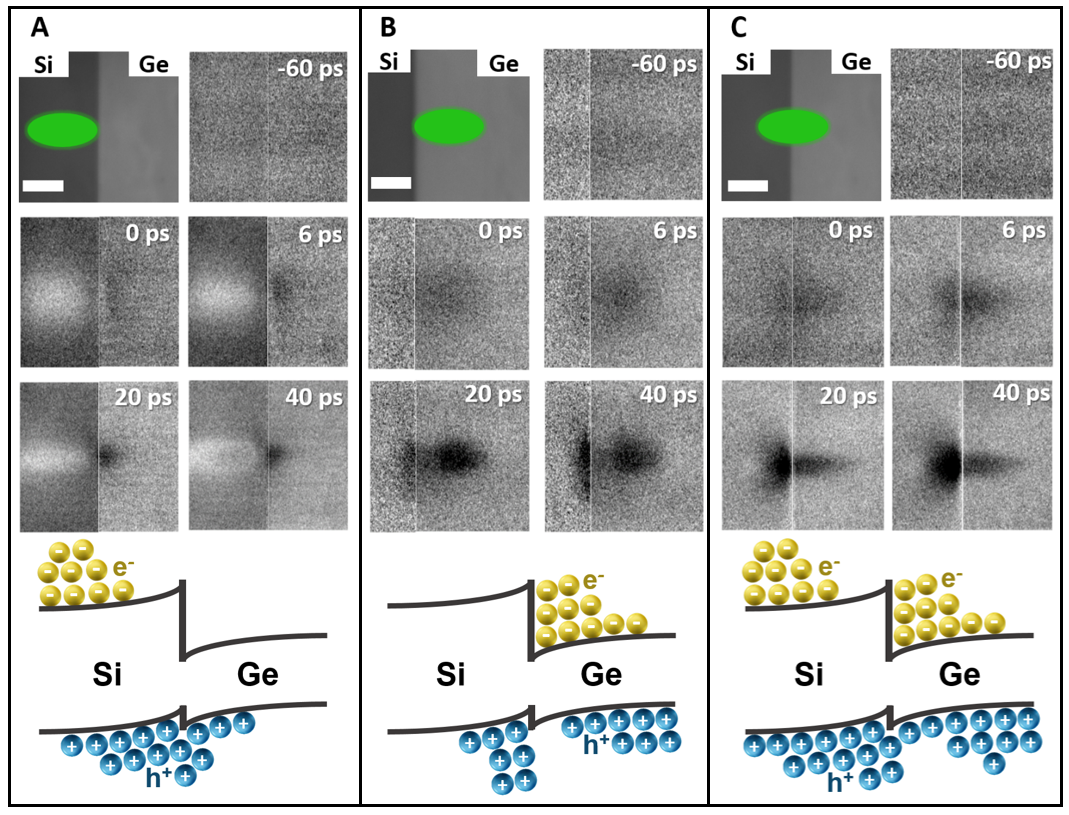}
\caption{\textbf{SUEM contrast images taken near the Si/Ge heterojunction.} (a) SUEM contrast images showing photocarriers excited on the Si side diffusing into the Ge side. The top left panel is a static SEM image of the junction, where the green ellipse indicates the location of the optical pump beam. The bottom panel shows a schematic of the photocarrier distribution as a result of the built-in potential and the band offsets of the heterojunction. (b) SUEM contrast images showing photocarriers excited on the Ge side diffusing into the Si side. (c) SUEM contrast images showing the dynamics of photocarriers excited right at the Si/Ge hetero-interface. Scale bar: 50 $\mu$m.} 
\label{fig:fig3}
\end{figure}

Next we use SUEM to directly visualize charge transfer across a Si/Ge heterojunction, as shown in Fig.~\ref{fig:fig3}. Figure~\ref{fig:fig3}a displays the scenario where the photocarriers are excited on the Si side (optical pump beam is indicated by the green ellipse in the static SEM image of the junction shown in the top left panel). As seen in the SUEM contrast images, initially after photoexcitation, bright contrast on the Si side is observed, which is consistent with the bulk Si result shown in Fig.~\ref{fig:fig2}. As the photoexcited carriers diffuse toward the junction, however, a contrast reversal occurs several micrometers away from the junction on the Si side. This effect is due to the built-in potential on the Si side (as shown in Fig.~\ref{fig:fig1}d) that attracts holes toward the junction while pushing the electrons away. The net effect is an accumulation of holes near the junction on the Si side, which leads to a reduced average energy of electrons and a lower SE yiled (dark contrast). Some of the hot holes are able to migrate across the valence band offset into the Ge side, leading to dark contrast on the Ge side as well. Figure~\ref{fig:fig3}b shows the opposite scenario, where the photocarriers are excited mostly on the Ge side. Due to the large conduction band offset (0.59\,eV, as shown in Fig.~\ref{fig:fig1}d), most of the photoexcited electrons cannot migrate into the Si side. Although the built-in potential on the Ge side tends to push holes away from the junction, some of the holes photoexcited very close to the junction can transfer into the Si side and be trapped by the potential valley due to the valence band offset. This effect can be clearly seen in the SUEM images taken at 20\,ps and 40\,ps, where the holes on the Si side diffuse more quickly parallel to the junction, but much more slowly perpendicular to the junction.

Figure~\ref{fig:fig3}c shows the case when the photocarriers are excited right at the junction. This case is analogous to the one studied by Najafi et al. in a Si p-n junction using SUEM~\cite{najafi2015four}, where the separation of photoexcited electrons and holes by the built-in potential was visualized. In our case, however, the presence of the heterojunction and the associated band offsets further complicates the photocarrier transport process. The built-in potential still tends to separate photoexcited electrons and holes on both sides of the junction, similar to that in a p-n homojunction. This is reflected in the emergence of a dark contrast on the Si side, where the built-in potential pushes electrons away from the junction while trapping the holes near the junction. At later delay times (20-40\,ps), bright contrast emerges on the Si side tens of micrometers away from the junction, indicating accumulation of electrons there as a result of the built-in potential separation effect. On the Ge side, this separation effect is not clearly seen because both electron and hole excitations lead to a dark contrast (Fig.~\ref{fig:fig2}b). However, a critical factor has so far been missing from the discussion. Given the high photon energy (2.4\,eV) associated with the optical pump pulse, photoexcited electrons and holes are in a ``hot'' state immediately after excitation with very high electronic temperatures, which can lead to diffusivities much higher than the near-equilibrium values. This initial superdiffusion behavior of photocarriers has been observed in previous SUEM measurements of many semiconductors~\cite{najafi2017super,liao2017photo,choudhry2023persistent,El-Zohry2019extra}. For example, in crystalline silicon, the initial diffusivity of photocarriers can be several orders of magnitude higher than the near-equilibrium values for up to 100\,ps~\cite{najafi2017super}, while in cubic boron arsenide, the superdiffusion process persists for over 200\,ps due to a hot phonon bottleneck effect~\cite{choudhry2023persistent}. In all previous studies, superdiffusion of hot carriers was investigated in uniform semiconductors, whereas the interaction of hot photocarriers with structures such as heterojunctions remains less explored. Given the ubiquity of heterojunctions in optoelectronic applications, it is thus important to understand how hot photocarriers behave near heterojunctions.

\subsection{Quantifying hot photocarrier diffusion near Si/Ge heterojunction}
\begin{figure}[!h]
\includegraphics[width=\textwidth]{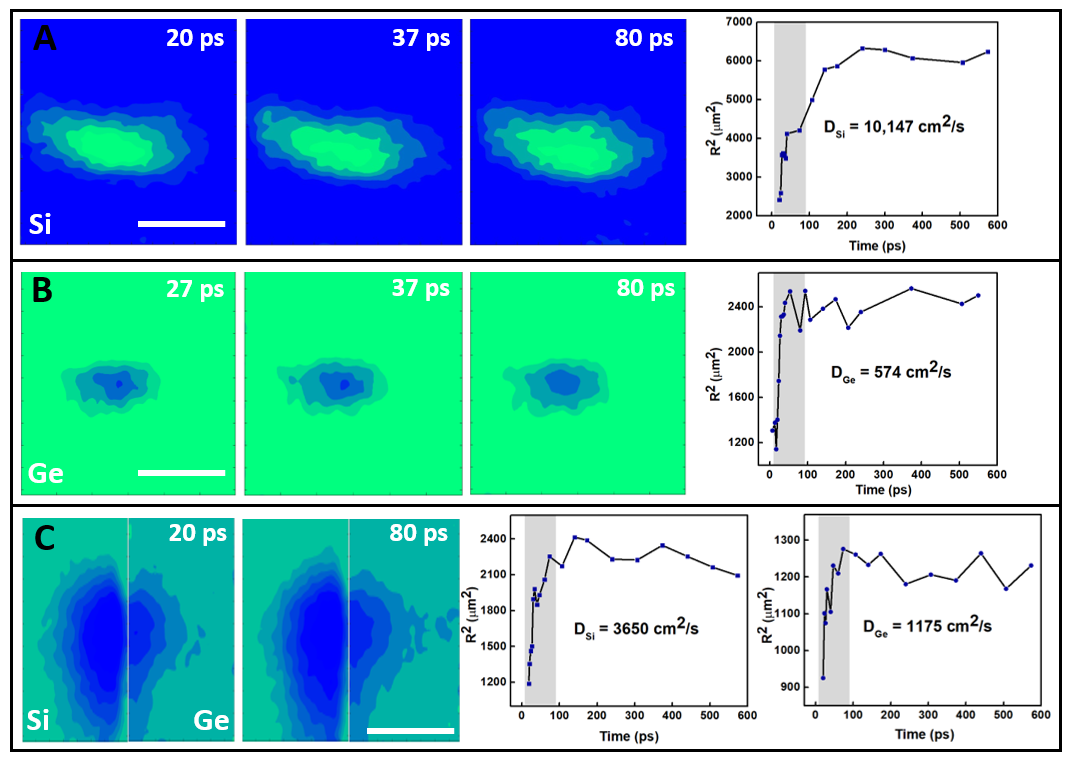}
\caption{\textbf{Diffusion of hot photocarriers excited near the Si/Ge heterojunction.} (a) The left three panels show contour plots of the SUEM contrast taken at bulk Si region far away from the junction as a function of delay time. The right panel displays the squared radius of the photocarrier distribution as a function of delay time. The shaded region indicates the superdiffusion regime where the hot photocarriers diffuse out rapidly immediately after photoexcitation. (b) Similar plots as (a) for SUEM contrast images taken in bulk Ge region far awary from the junction. (c) Similar plots as (a) for SUEM contrast images when photocarriers are excited at the heterojunction. The right two plots show the squared radius of the photocarrier distribution as a function of delay time for the Si and Ge side of the junction, respectively. Scale bar: 50 $\mu$m.} 
\label{fig:fig4}
\end{figure}

In this section, we quantify the influence of the Si/Ge heterojunction on recombination and diffusion processes of hot photocarriers immediately after excitation from the SUEM contrast images. The fitted intensity of the photocarrier contrast near the Si/Ge heterojunction as a function of delay time is shown in Fig.~S3c, where the recombination time in both Si and Ge regions are much longer than those measured in bulk regions far away from the junction. This is a consequence of the charge-separation effect of the junction built-in potential, which reduces the spatial overlap of electrons and holes and, thus, the probability of radiative recombination. Stages of photocarrier diffusion in uniform bulk Si have been visualized and analyzed by SUEM in a previous work~\cite{najafi2017super}. Here, our observation in the bulk Si region (as shown in Fig.~\ref{fig:fig4}a) is similar to the previous report. The first three panels in Fig.~\ref{fig:fig4}a show contour plots with constant SUEM contrast intensity as a function of delay time in the bulk Si region. Tracking the spatial-temporal evolution of the contours can provide quantitative information about the photocarrier diffusion process. Alternatively, the SUEM contrast images can be fitted to 2D Gaussian functions with a time-dependent radius, which is shown in the fourth panel of Fig.~\ref{fig:fig4}a. The slope of the squared radius of the Gaussian distribution as a function of delay time represents the diffusivity at a given moment. It can be clearly seen that photoexcited carriers diffuse out rapidly right after photoexcitation (with a diffusivity of about 10,000\,cm$^2$/s) for a time period of about 100 ps, before slowing down and eventually approaching the near-equilibrium ambipolar diffusivity value of about 20\,cm$^2$/s. Our result in the bulk Si region agrees quantitatively with the previous SUEM experiment~\cite{najafi2017super}. In the bulk Ge region (Fig.~\ref{fig:fig4}b), we observed a similar hot photocarrier diffusion regime, nevertheless with a much lower hot carrier diffusivity of 512\,\,cm$^2$/s compared to that in Si.  Although both electrons and holes in Ge have higher near-equilibrium mobilities (and thus diffusivities) than those in Si, a few factors can possibly lead to much lower hot carrier diffusion. First, Ge has a smaller band gap (0.66\,eV) compared to Si (1.11\,eV), suggesting electrons and holes can be excited by the optical pump (2.4\,eV) to higher-energy bands with heavier mass and lower mobility. Second, surface bands are observed to play a significant role in pinning the Fermi level on Ge surface regardless of doping concentration~\cite{gobeli1964photoelectric}, suggesting a higher surface defect concentration that can scatter the photocarriers and slow down their diffusion. This view is also consistent with the short recombination time measured in the bulk Ge region (Fig.~S3b).

Figure~\ref{fig:fig4}c shows the photocarrier diffusion process near the Si/Ge heterojunction when the photocarriers are excited right at the interface. Influenced by the built-in potential near the junction, photocarrier diffusion on both sides becomes more anisotropic than the bulk regions. Moreover, the hot carrier diffusivities near the heterojunction are significantly modified. On the Si side, mainly the diffusion of the photoexcited holes is imaged due to the charge separation effect of the built-in potential as discussed in the previous section. The hot carrier diffusion on the Si side is slowed down to a diffusivity of 3650\,cm$^2$/s. This can be understood as a combination of the built-in potential that tends to trap holes near the junction and the injection of slow holes from the Ge side (while only a relatively small fraction of electrons can be injected from the Ge side due to the larger conduction band offset). In contrast, on the Ge side, the hot carrier diffusivity is increased to about twice the bulk value (1175\,cm$^2$/s). This effect can be mainly attributed to the injection of hot electrons from the Si side, which are further accelerated by the conduction band offset. In this case, the built-in potential and the band offsets in the Si/Ge heterojunction act together to homogenize the difference in the hot carrier diffusion processes in bulk Si and Ge regions. This direct visualization of hot photocarrier dynamics near a heterojunction enabled by the development of SUEM can provide an experimental basis to benchmark photocarrier transport theories and simulations, which can lead to better understanding and design of optoelectronic devices based on semiconductor heterojunctions.

\subsection{Conclusion}
In summary, we use SUEM to directly visualize photocarrier dynamics near a semiconductor heterojunction in space and time for the first time. The unique combination of space and time resolutions and surface sensitivity of SUEM allows us to explore the interaction between photoexcited charge carriers and the junction built-in potential, band offsets, and the surface effects, assisted by other characterization techniques that provide comprehensive parameters of the Si/Ge heterojunction under study. In particular, we observe the impact of the heterojunction on the initial diffusion of hot photocarriers excited right at the interface, which can have significant implications for hot-carrier based photovoltaic, photocatalysis, and photosensing devices. Our study also showcases SUEM as an emerging technique with the potential to image charge-transport processes in realistic devices.

\section{Methods}
\subsection{Sample Preparation}
(001) Germanium was bonded to (001) silicon using EVG\textsuperscript{\textregistered} ComBond\textsuperscript{\textregistered} equipment under high vacuum (~10$^{-8}$mtorr) and room temperature. This process is similar to other reported surface activated bonding (SAB) or in situ sputtering techniques, in which Ge and Si surfaces are sputtered with an Ar ion
beam (300\,eV) at a shallow angle of 45 degrees to remove unwanted native surface oxides prior to bonding~\cite{mu2019high,xu2019direct}. The post Ar-beam treated samples are placed face-to-face and pressure is applied to initiate the bond. The Si crystal (thickness of 700 $\mu$m, 1 $\times$ 1.5\,cm) is p-type with a doping concentration of 10$^{16}$\,cm$^{-3}$. The Ge crystal (thickness of 700 $\mu$m, 1 $\times$ 1.5\,cm) is n-type with the same doping concentration of 10$^{16}$\,cm$^{-3}$. 

\subsection{Sample Characterization}

Structural characterization for Si/Ge heterojunction was conducted using electron microscopy techniques including high resolution scanning transmission electron microscopy (STEM) and energy dispersive X-ray spectroscopy (EDX). An FEI Nova 600 Nanolab Dual Beam SEM/FIB was used to prepare cross section TEM samples roughly 100\,nm thick using a Ga source and then transferred to a TEM grid using a standard lift-out procedure. High-resolution transmission
electron microscopy (HRSTEM) of the interface was taken with a Cs-corrected JEOL GrandARM at 300\,kV. Asylum Research Jupiter Atomic Force Microscopy (AFM) was used in scanning Kelvin probe mode (SKPM) to measure the surface potential on the Si-Ge cross-section sample with a scan angle of 90 degrees using a silicon tip with Ti/Pt coating. 

In order to determine the energy band offsets at Si-Ge interface, Kraut’s method was followed in X-ray photoelectron spectroscopy (XPS) analysis \cite{kraut1980precise,kraut1983semiconductor}. Thermo Fisher Escalab Xi+ XPS Microprobe with a monochromated Al K$\alpha$ X-ray source (beam energy of 1486.7\,eV) was used to measure the binding energy associated with core levels (CLs) and valence band maxima (VBM). The XPS spectra were collected from three locations on the Si/Ge cross section: (1) the bulk Si region to measure the CL binding energy of Si 2$p$ and the VBM; (2) the bulk Ge region to measure the CL binding energy of Ge 3$d$ and the VBM; (3) the Si-Ge heterojunction region to measure the CL binding energy of Si 2$p$ and Ge 3$d$ at the interface. In order to neutralize positive charge accumulation on the surface and overcome the shift in binding energy, charge compensation by an electron flood source was used. Besides, all the binding energy values were corrected by shifting the carbon 1$s$ CL peak to 285\,eV.

\subsection{Scanning Ultrafast Electron Microscopy}
 
The detailed description of the SUEM setup was provided in our previous publications ~\cite{kim2021transient,choudhry2023persistent}. Briefly, a fundamental infrared (IR) laser (Clark MXR IMPULSE) operating at a central wavelength of 1030\,nm with a pulse duration of 150\,fs is directed to frequency-doubling crystals to create the visible pump beam (wavelength 515\,nm) and the ultraviolet (UV) photoelectron excitation beam (wavelength 257\,nm).  The UV excitation beam is directed through the column of an SEM (Thermo Fisher Quanta 650 FEG) and onto the apex of a cooled Schottky field emission gun (ZrO$_2$-coated tungsten) to generate electron pulses with picosecond durations via the photoelectric effect. The photo-generated electron pulses are accelerated inside the SEM column to 30\,keV kinetic energy. The visible pump beam is directed inside the microscope to initiate the excitation of the sample. The diameter of the optical pump beam used in this study is around 30\,$\mu$m. The distance traveled by the visible pump beam is adjusted by a computer-controlled mechanical delay stage (Newport DL600, delay time range -0.7 to 3.3\,ns). All experiments reported here were conducted at a laser repetition rate of 5\,MHz, an electron probe beam current of 30 to 40\,pA and a visible pump beam fluence of 20\,$\mu$J/cm$^2$. This particular optical fluence was chosen to achieve the highest signal-to-noise ratio while avoiding permanent optical damage of the sample at higher fluences. Considering the near 45$^{\circ}$ incident angle of the optical pump beam, the excited photocarrier concentration is estimated to be around $5 \times 10^{17}$\,cm$^{-3}$ in Si and $1 \times 10^{19}$\,cm$^{-3}$ in Ge based on their documented optical properties. The probe beam current corresponds to 30 to 40 electrons per pulse and a pulse width of a few ps~\cite{choudhry2023persistent}. The images were acquired using a dwell time of 300\,ns per pixel and an integration of 256 frames. The samples were mounted on an SEM cross-section stub to measure the interface as received after bonding without polishing or other surface treatment.

\bibliography{references.bib}

\begin{acknowledgments}
This work is based on research partially supported by the U.S. Air Force Office of Scientific Research under award number FA9550-22-1-0468 (for studying hot photocarrier dynamics) and the U.S. Army Research Office under award number W911NF2310188 (for the development of SUEM). Y.Q. also acknowledges the support from the Graduate Traineeship Program of the NSF Quantum Foundry via the Q-AMASE-i program under award number DMR-1906325 at the University of California, Santa Barbara (UCSB).A portion of this work was performed in UCSB Nanofabrication Facility, an open-access laboratory. XPS and SKPM measurements were conducted in the UCSB Microscopy and Microanalysis Facility, which is part of the UCSB Materials Research Laboratory Shared Experimental Facilities supported by the MRSEC Program of the NSF under Award Number DMR-2308708; a member of the NSF-funded Materials Research Facilities Network (www.mrfn.org).

\end{acknowledgments}

\end{document}



\title{Supplementary Materials: Imaging Hot Photocarrier Transfer across a Semiconductor Heterojunction with Ultrafast Electron Microscopy} 

\author{Basamat Shaheen}
\affiliation{Department of Mechanical Engineering, University of California, Santa Barbara, CA 93106, USA}

\author{Kenny Huynh}
\affiliation{Department of Materials Science and Engineering, University of California, Los Angeles, CA 90095, USA}

\author{Yujie Quan}
\affiliation{Department of Mechanical Engineering, University of California, Santa Barbara, CA 93106, USA}

\author{Usama Choudhry}
\affiliation{Department of Mechanical Engineering, University of California, Santa Barbara, CA 93106, USA}

\author{Ryan Gnabasik}
\affiliation{Department of Mechanical Engineering, University of California, Santa Barbara, CA 93106, USA}

\author{Zeyu Xiang}
\affiliation{Department of Mechanical Engineering, University of California, Santa Barbara, CA 93106, USA}

\author{Mark Goorsky}
\email{goorsky@seas.ucla.edu}
\affiliation{Department of Materials Science and Engineering, University of California, Los Angeles, CA 90095, USA}

\author{Bolin Liao}
\email{bliao@ucsb.edu} \affiliation{Department of Mechanical Engineering, University of California, Santa Barbara, CA 93106, USA}

\date{\today}

\maketitle
\renewcommand{\thefigure}{S\arabic{figure}}
\renewcommand{\thetable}{S\arabic{table}}


\section*{Supplementary Text: }

\section{Analysis of the Heterojunction Potentials based on XPS}
\begin{figure}[!h]
\includegraphics[width=\textwidth]{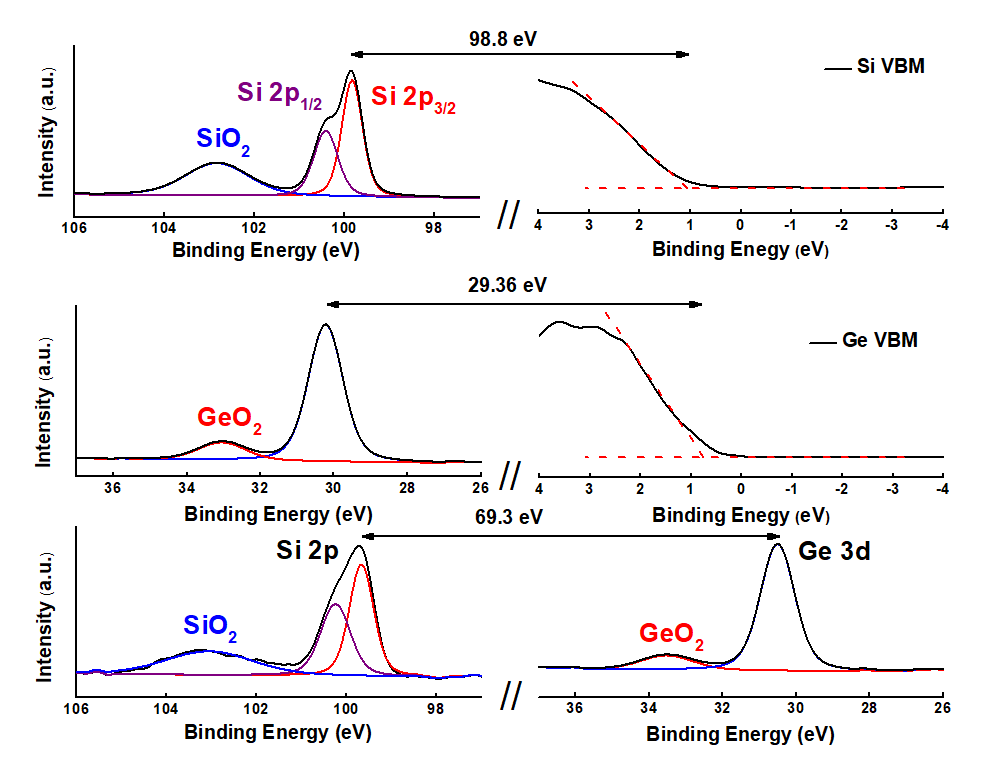}
\caption{\textbf{XPS spectra measured across the Si/Ge heterojunction.} The top panel shows the XPS spectrum measured in the Si bulk region, where the peak corresponding to the Si 2$p$ level and the continuum corresponding to the valence band maximum (VBM) are labeled. The middle panel shows the XPS spectrum measured in the Ge bulk region, where the peak corresponding to the Ge 3$d$ level and the continuum corresponding to the VBM are labeled. The lower panel shows the XPS spectrum measured at the heterojunction interface, where the peaks corresponding to the Si 2$p$ level and the Ge 3$d$ level are labeled.} 
\label{sfig:XPS}
\end{figure}

X-ray photoemission spectroscopy (XPS) is a widely used surface-sensitive technique to determine band offsets and built-in potentials across heterojunctions. XPS directly measures the binding energy, which is the energy difference between the Fermi level and a selected core level. Valence band offsets at a heterojunction can be determined by Kraut's method~\cite{kraut1980precise,kraut1983semiconductor}, where the valence band offset can be determined by conducting XPS measurements in bulk regions on both sides and at the interface. Raw XPS data measured from our Si/Ge heterojunction sample is given in Fig.~\ref{sfig:XPS}. The valence band offset is given by~\cite{kraut1980precise,chambers2020introductory}:
\begin{equation}
    \Delta E_V = (E_{CL1}-E_{V1})_{\mathrm{bulk1}} - (E_{CL2}-E_{V2})_{\mathrm{bulk2}} - (E_{CL1}-E_{CL2})_{\mathrm{interface}}, 
\end{equation}
where $\Delta E_V$ is the valance band offset, $(E_{CL1}-E_{V1})_{\mathrm{bulk1}}$ and $(E_{CL2}-E_{V2})_{\mathrm{bulk2}}$ are binding energy differences between a chosen core level and the valence band maximum (VBM) measured in bulk regions on both sides of the heterojunction, respectively, and $(E_{CL1}-E_{CL2})_{\mathrm{interface}}$ is the binding energy difference of the two core levels measured at the heterojunction interface. In our case, Si 2$p$ and Ge 3$d$ are chosen as the core levels. The corresponding binding energy differences are labeled in Fig.~\ref{sfig:XPS}. Based on the XPS data, the valence band offset is calculated to be 0.14\,eV. Then, the conduction band offset $\Delta E_C$ can be determined by~\cite{chambers2020introductory}:
\begin{equation}
    \Delta E_C = \Delta E_V + (E_{g1}-E_{g2}),
\end{equation}
where $E_{g1}$ and $E_{g2}$ are the bandgaps of the two materials forming the junction. Given the Si bandgap of 1.11\,eV and the Ge bandgap of 0.66\,eV, the conduction band offset $\Delta E_C$ is calculated to be 0.59\,eV. Additionally, given the VBM binding energy difference measured in the bulk Si and Ge regions ($\sim$0.36\,eV), the total built-in potential across the heterojunction can be estimated to be 0.5\,eV. The band diagram based on this analysis is given in Fig.~1d in the main text. 

\clearpage

\section*{Supplementary Figures}

\clearpage

\begin{figure}[!htb]
\includegraphics[width=0.5\textwidth]{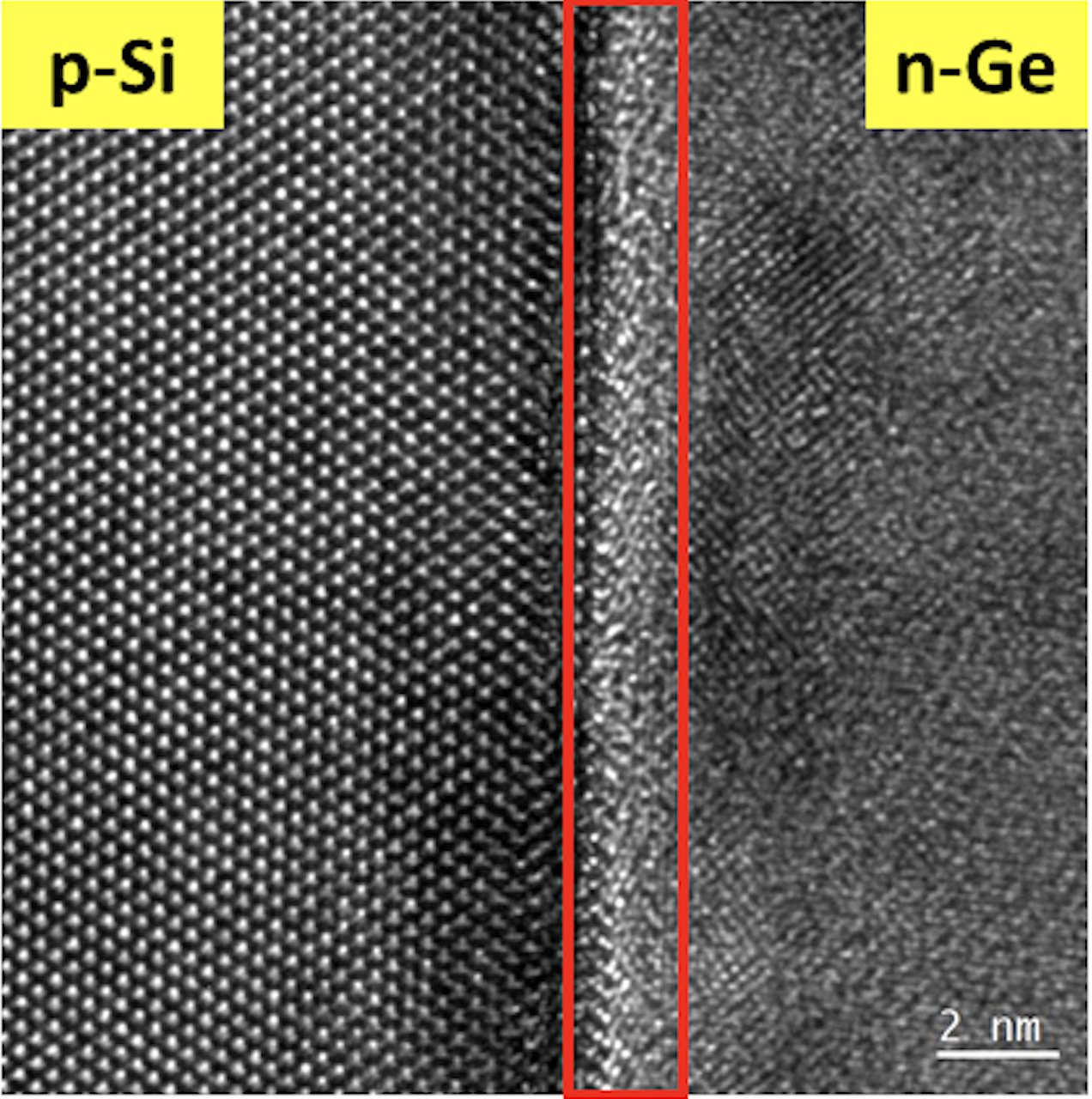}
\caption{TEM image of the Si/Ge heterojunction after wafer bonding.} 
\label{fig:figS2}
\end{figure}

\clearpage

\begin{figure}[!htb]
\includegraphics[width=\textwidth]{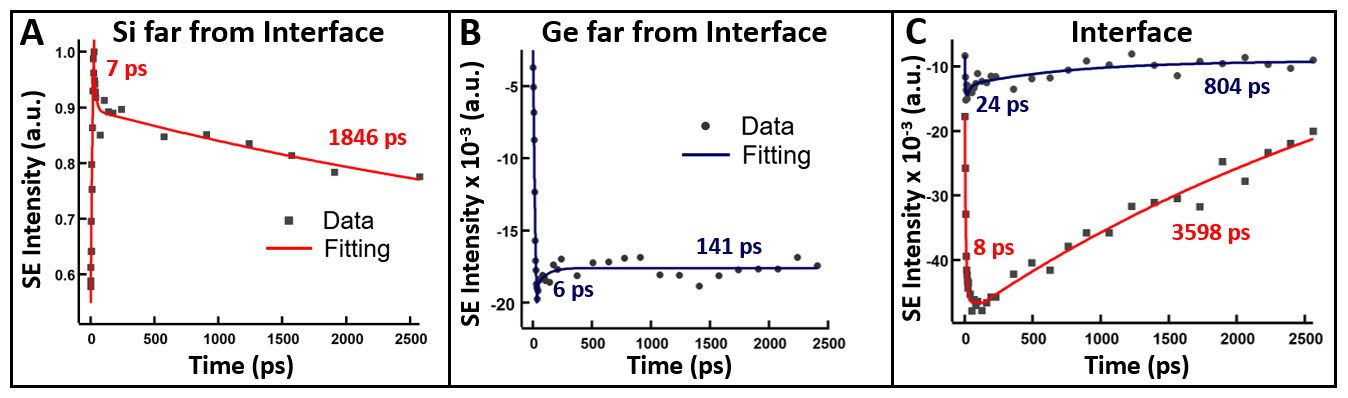}
\caption{The intensity of the SUEM contrasts as a function of delay time taken in (a) the bulk Si region far from the junction, (b) the bulk Ge region far from the junction, and (c) the Si side (red line) and Ge side (black line) of the heterojunction. The intensities are extracted by fitting the raw SUEM contrasts to 2D Gaussian functions with time-dependent intensity and radius. } 
\label{fig:figS3}
\end{figure}

\clearpage
\textbf{Supplementary References}
\bibliography{references.bib}